# Fabrication and characterization of Er,Nd codoped $Y_2O_3$ Transparent Ceramic: A dual mode Photo-luminescence emitter

**Pratik Deshmukh[a,*], S. Satapathy[a,c*], Anju Ahlawat[a], M K Tiwari[b] and A. K. Karnal[a,c]**

[a] *Laser Materials Section, Raja Ramanna Centre for Advanced Technology, Indore 452013, India*

[b] *Synchrotron Utilization Section, Raja Ramanna Centre for Advanced Technology, Indore 452013, India*

[c] *Homi Bhabha National Institute, BARC, Mumbai, India.*

**Corresponding author**

[*] **E- mail:** ppdeshmukh@rrcat.gov.in; srinu73@cat.ernet.in

Phone: 91 731 2442905/2920/2010

Fax: 91 731 2442900

**Abstract:**

The transparent Er, Nd co-doped $Y_2O_3$ ceramics with transparency ~78% (in 500-2000 nm range without Fresnel's correction) were fabricated successfully. It involved nanoparticle synthesis by coprecipitation method and sintering of pellets under high vacuum condition. The crystalline phase, particle size and element composition were confirmed by X-ray diffraction, scanning electron microscope, energy dispersive X-ray fluorescence techniques, respectively. The upconversion luminescence mechanisms involving energy transfer and non-radiative relaxation were analyzed. It is expected that, the result evolved from this study will provide better understanding of upconversion mechanism involved in Er, Nd codoped host material. The emission at both 563 nm ($Er^{3+}:{}^4S_{3/2} \rightarrow {}^4I_{15/2}$) and 1064 nm ($Nd^{3+}:{}^4F_{3/2} \rightarrow {}^4I_{9/2}$) on 822 nm ($Nd^{3+}:{}^4I_{9/2} \rightarrow {}^4F_{5/2}$) excitation proves potential of the ceramic material for dual mode efficient emitter.

1. **Introduction**

The transparent ceramics application as a laser gain medium has increased after the evolution of thin disk laser concept. This concept is based on a laser design for the diode pumped solid state lasers, which leads to achieve high efficient laser oscillations in compact device [1-2]. The transparent ceramic of various materials such as $Y_2O_3$, YAG, $Sc_2O_3$ and $La_2O_3$ were already well fabricated [3-6]. The high transmittance in wide spectral range (i.e. from visible to Infra-red) proves transparent ceramic's potential for upconversion (UC) laser application [7].

Upconversion is anti-stoke emission type process referred as nonlinear optical process where sequential absorption of two or more photons leads to emission of a single photon having shorter wavelength than that of incident one [8]. It has advantages over other nonlinear optical process such as second harmonic generations because of no strict requirement of coherent excitation source, phase matching condition and operation at comparatively low incident power [9-10].

Amongst the various dopant in solid state materials, most efficient UC emission occurs with rare earth ions [8]. However, Laporte-forbidden transition (i.e. 4f-4f) limits the absorption of incident light. The increase in dopant concentration increases the absorption up-to a limit over which cross-relaxation dominates. The use of sensitizer in lanthanide doped phosphor is often preferred, which is a strong absorbing ion and transfer the energy to activator efficiently.

The $Er^{3+}$ based upconversion phosphor is widely used with $Yb^{3+}$, which acts as sensitizer for 980 nm excitation wavelength. The energy difference between $^2F_{7/2}$ to $^2F_{5/2}$ transition of $Yb^{3+}$ matched well with several f-f transitions in $Er^{3+}$. Thus, allowing resonant energy transition between them [8,11]. Similarly, $Nd^{3+}$ doped material mostly utilized for efficient laser oscillator. In addition, the absorption cross-section of $Nd^{3+}$ around 808 nm is much larger (about 10 times) compared to $Yb^{3+}$ ions at 980 nm [12]. $Nd^{3+}$ and $Er^{3+}$ codoped nanophosphor for bioimaging in dual mode (i.e. emission in visible and IR region simultaneously) on 800 nm excitation have been reported by Li et al [13]. Thus, its use as sensitizer not only improve efficiency of upconversion process but also leads to dual mode emitter as described in fig. 1.

In this paper, Nd and Nd, Er codoped $Y_2O_3$ transparent ceramics were fabricated by utilizing vacuum sintering for densification of nanoparticles synthesized by co-precipitation method. Subsequently, the photoluminescence characterization has been studied. The upconversion mechanism leading to intense emission at 563 nm (Er:$^4S_{3/2} \rightarrow {}^4I_{15/2}$) on 822 nm (Nd:$^4I_{9/2} \rightarrow {}^4F_{5/2}$) excitation have been investigated. To the best of our knowledge, Er, Nd co-doped $Y_2O_3$ transparent ceramics fabrication and photoluminescence characterization have not been reported yet.

## 2. Materials and methods

The nanoparticles synthesis with uniform morphology and narrow particle size distribution is considered as an important parameter involved in the fabrication of transparent ceramics. Here, the co-precipitation method was used to synthesize nano powder of Er(x at%),Nd(1at%): $Y_2O_3$ (where, x= 0 and 1) with additional $Zr^{4+}$ and $La^{3+}$ dopants [14-15]. These additional dopants ($Zr^{4+}$(2at%) and $La^{3+}$(1at%)) acts as sintering additives. Mother solution was prepared by dissolving $La_2O_3$, $Nd_2O_3$, $Er_2O_3$, $Y_2O_3$ (purity 99.99%) and $ZrOCl_2 \cdot 8H_2O$ (purity 98%) (0.2M) in dilute nitric acid (1.5M). The precipitant ammonium solution was added gradually in nitrate solution under vigorous stirring. Ammonium sulfate solution was used as surfactant, which is added in dispersant solution after an hour of aging. Washing of precipitate was carried out for 5 h using methanol as washing solvent. Further, the washed precipitate was filtered, dried (at 80°C) and calcined at 900°C to get desired oxide powder. This oxide powder was pelletized using uniaxial press and finally vacuum sintered at 1750 °C for 5 h.

Rigaku X-ray diffractometer, Zeiss field emission scanning microscope (FESEM) and Malvern Zetastar 90 (model no-ZEN 3690) particle size analyzer were utilized for the characterization of nanoparticles. Chemical composition of the sintered pellets was analyzed using energy dispersive X-ray fluorescence (EDXRF) measurement at BL-16 beamline of Indus-2 synchrotron radiation facility [16]. The quantitative analysis was carried out by Fundamental parameter method using CATXRF program [17]. The sintered specimen was double-side polished (1.0 mm thick) and its optical spectra were measured on a V-670 transmission spectrophotometer (JASCO Corporation). The upconversion and downshifting luminescence spectra were obtained with a FLS920-s fluorescence spectrometer (Edinburgh Instruments Ltd.) upon excitation at 822 nm. All characterizations were performed at room temperature.

3. **Result and discussion**

The XRD patterns of prepared $Y_2O_3$: $Er^{3+}$(1at%), $Nd^{3+}$(1at%) and $Y_2O_3$:$Nd^{3+}$(1at%) nanopowders along with JCPDS Card No. 41-1105 are shown in fig. 2. The characteristic diffraction peaks of as-prepared samples matched well with the standard cubic $Y_2O_3$ (JCPDS Card No. 41-1105) phase with space group Ia-3. Also, no additional peak is found in the pattern, indicating that dopants concentrations are under solubility limit.

Figure 3(a) and 3(b), respectively, shows morphology of $Y_2O_3$: $Nd^{3+}$(1 at%) and $Y_2O_3$: $Er^{3+}$(1 at%), $Nd^{3+}$(1 at%) powders after calcination. The particles are physically overlapped and are having nearly spherical morphology. The particle size distribution of the samples, measured using dynamic light scattering method is shown in figure 3(c). The narrow range (~10-80 nm) of particle size distribution shows the suitability of nanoparticles for transparent ceramic fabrication.

Figure 4(a) and 4(b) shows the characteristic XRF signals of Nd, La, Zr and Y in $Y_2O_3$:$Nd^{3+}$(1at%) and additional Er signal in $Y_2O_3$:$Er^{3+}$(1at%),$Nd^{3+}$(1at%). It consists of characteristic peaks with superimposed spectral artifacts and well-defined background. The Y fluorescent lines ($K_\alpha$ and $K_\beta$) are well distinguishable for 19.7 keV excitation energy. Several small peaks in the spectra confirm the presence of various dopant ions Zr, La, Nd and Er in host $Y_2O_3$. The elemental composition of both samples $Y_2O_3$:$Nd^{3+}$(1at%) and $Y_2O_3$:$Er^{3+}$(1at%),$Nd^{3+}$(1at%) were determined using XRF measurement and are shown in table1. It matches quit well with the nominal compositions within the predicted error range of XRF technique. The statistical error in peak intensity determination is the dominant error, which varies with element concentration.

The optical transmittance of $Y_2O_3$: $Er^{3+}$(1at%), $Nd^{3+}$(1at%) sintered pellet of 1 mm thickness is shown in fig. 5. the Light transmittance is the important parameter for estimating the optical properties of transparent ceramics. The transmittance reached ~78% at 564 nm and almost remains constant at higher wavelengths except at characteristic absorption peaks of dopants. The spectrum consists of various absorption bands originating from ground state of both dopants $Nd^{3+}$ and $Er^{3+}$. Most of the prominent absorption bands are designated in spectrum. The band centered at 592, 743, 822 and 896 nm are attributed to transition from ground state $^4I_{9/2}$ to excited states $^2G_{7/2}+^4G_{5/2}$, $^4S_{3/2}+^4F_{7/2}$, $^2H_{9/2}+^4F_{5/2}$ and $^4F_{3/2}$ of $Nd^{3+}$, respectively [18]. Also, some well separated absorption peaks of $Er^{3+}$ are observed at 380, 523 and 1536 nm due to ground state ($^4I_{15/2}$) to $^4G_{11/2}$, $^2H_{11/2}$, and $^4I_{13/2}$ transitions respectively [19-20].

Figure 6(a) shows the infrared excitation spectra of both ($Y_2O_3$:$Er^{3+}$(1at%)$Nd^{3+}$(1at%)and $Y_2O_3$:$Nd^{3+}$(1at%)) transparent ceramics. The recorded spectra, on monitoring emission at 1064 nm, shows several prominent excitation peaks around 822 and 895 nm corresponding to transition from ground state $^4I_{9/2}$ to excited states $^4F_{5/2}$+ $^2H_{9/2}$ and $^4F_{3/2}$ respectively [18]. The peak positions in excitation spectra are same irrespective of dopants in $Y_2O_3$ transparent ceramic but relatively lower intensity peaks observed for co-doped sample. This indicates the energy transfer is possible between the corresponding dopants.

Figure 6(b) shows the PL emission spectra, under 822 nm excitation, of $Y_2O_3$:$Er^{3+}$(1at%)$Nd^{3+}$(1at%) and $Y_2O_3$:$Nd^{3+}$(1at%) samples in 420 to 1700 nm range. In this measured range for $Y_2O_3$: $Nd^{3+}$(1at%), three emission bands 870-985 nm, 1040-1196 nm and 1306-1469 nm corresponding to transition from $^4F_{3/2}$ to $^4I_{9/2}$, $^4I_{11/2}$ and $^4I_{13/2}$ energy level of $Nd^{3+}$, respectively, were observed. In the case of co-doped sample ($Y_2O_3$:$Er^{3+}$(1at%)$Nd^{3+}$(1at%)), additional emission peaks observed at 500, 548, 563, 660 and 1535 nm, which corresponds to transitions from excited levels $^4F_{7/2}$, $^2H_{11/2}$, $^4S_{3/2}$, $^4F_{9/2}$ and $^4I_{13/2}$ to ground level $^4I_{15/2}$ of $Er^{3+}$ respectively. It is worth noting that there is a significant change in emission intensity of $Nd^{3+}$ band in Er, Nd co-doped $Y_2O_3$ as compared to the single $Nd^{3+}$ doped $Y_2O_3$. The emission peak at 563 nm is prominent in Er, Nd co-doped sample while no prominent visible emission is observed in Nd doped sample.

The upconversion emission mechanism is explained using the energy level diagram of both $Nd^{3+}$ and $Er^{3+}$ as shown in fig. 7. The absorption of 822 nm incident radiation excites $Nd^{3+}$ ion via ground state absorption to $^4F_{5/2}$ + $^2H_{9/2}$. Further, emissions from $Nd^{3+}$ ion occurs when it decays non-radiatively to metastable $^4F_{3/2}$ from which it relaxed radiatively to the corresponding lower energy levels [21]. In the case of co-doped sample, $Nd^{3+}$ ion transfer the part of absorbed energy to $Er^{3+}$ via energy transfer (ET) process: $^4F_{3/2}$ ($Nd^{3+}$) + $^4I_{15/2}$($Er^{3+}$)→ $^4I_{9/2}$ ($Nd^{3+}$)+ $^4I_{11/2}$($Er^{3+}$) [22-23].This leads to decrease in intensity in emission peaks corresponding to $Nd^{3+}$ ion. Meanwhile, the lower energy level $^4I_{13/2}$ of $Er^{3+}$ is also populated because of nonradiative relaxation from upper $^4I_{11/2}$ level [16]. The electrons in these two energy levels of $Er^{3+}$ gets further excited to the higher excited states $^4F_{3/2}$ and $^2H_{11/2}$ via two excited state absorption (ESA) processes: $^4I_{11/2}$($Er^{3+}$) + $E_{photon}$ → $^4F_{3/2}$($Er^{3+}$) and $^4I_{11/2}$($Er^{3+}$) + $E_{photon}$ → $^2H_{11/2}$($Er^{3+}$), respectively. Finally, these populated $Er^{3+}$ level relaxes dominantly through multi-phonon relaxation (MPR) by non-radiative transition to the next lower levels ($^2H_{11/2}$, $^4S_{3/2}$ and $^4F_{9/2}$), which subsequently results in visible emission on radiative transitions from these states to the ground state.

Figure 8 shows Commission Internationale de l'Eclairage (CIE) diagram for $Y_2O_3$:$Er^{3+}$(1at%)$Nd^{3+}$(1at%) sample. The calculated colour coordinates (x, y) are (0.355, 0.637). These values fall in yellowish-green region of CIE diagram.

Besides upconversion, the downconversion luminescence in both samples have been investigated. Figure 9 shows the excitation spectra measured in 300-400 nm range by monitoring emission at 1064 nm derived from $^4F_{3/2} \to {}^4I_{11/2}$ transition of $Nd^{3+}$ and the emission spectra using 357 nm as excitation wavelength derived from $^4I_{9/2} \to {}^4D_{3/2}$ transition of $Nd^{3+}$. The near IR down-conversion material suitable for c-Si solar cell must be able to convert near UV (300-400 nm) part of solar spectrum to ~1000-1100 (band gap 1.12 eV) nm photons. The result shows, Nd:$Y_2O_3$ transparent ceramic is an efficient down-conversion material and hence had application for enhancing the efficiency of C-Si solar cell.

## 4. Conclusions

The $Er^{3+}$ and $Nd^{3+}$ co-doped $Y_2O_3$ transparent ceramics fabrication and infrared to visible frequency upconversion and near infrared downshifting has been investigated for the first time. The up-conversion was because of the energy transfer from $Nd^{3+}$ to $Er^{3+}$ ion, which are coexisted in $Y_2O_3$ host. Additional dopants $Zr^{4+}$ and $La^{3+}$ were used to increase sinter-ability of transparent ceramic. The phase formation, particle size and element composition were estimated using XRD, SEM and EDXRF measurement, respectively. The intrinsic characteristics of host $Y_2O_3$ involving its low photon energy, high thermal conductivity, broad transparency range, chemical and mechanical stability makes $Er^{3+}$ and $Nd^{3+}$ co-doped $Y_2O_3$ as promising dual mode efficient emitter. The obtained result in downconversion luminescence also put forward the possible use of the present transparent ceramic as spectral convertor in photovoltaics.

**Acknowledgements**

We thank Dr. Gurvinderjit Singh, Mr. S. K. Pathak (Laser Materials Section), Mr. M. P. Kamath and Mr. Pawan Kumar (Optical Design and Development Laboratory, Advanced Lasers & Optics Division) RRCAT, Indore for sintering and polishing of the pellets.

**Table 1: Elemental Composition of the Particles using EDXRF**

| Sr No | Sample | Elements | Concentration (%) |
|---|---|---|---|
| 1. | $Y_2O_3$: Nd,Zr,La | Nd | 0.77 ± 0.1 |
|  |  | Y | 96.84 ± 1 |
|  |  | La | 1.00 ± 0.1 |
|  |  | Zr | 1.39 ± 0.5 |
| 2. | $Y_2O_3$:Er,Nd,Zr,La | Er | 1.25 ± 0.3 |
|  |  | Nd | 0.72 ± 0.1 |
|  |  | Y | 95.42 ± 1 |
|  |  | La | 1.00 ± 0.1 |
|  |  | Zr | 1.61 ± 0.5 |

**Figure Captions:**

**Fig. 1.** General strategy to achieve the upconversion and downshifting luminescence with Er/Nd codoped transparent ceramic.

**Fig. 2.** X-ray diffraction pattern of $Y_2O_3$:$Er^{3+}$(1at%), $Nd^{3+}$(1at%) and $Y_2O_3$:$Nd^{3+}$(1at%) nanopowder.

**Fig. 3.** FESEM image of (a) $Y_2O_3$:$Er^{3+}$(1at%), $Nd^{3+}$(1at%) and (b) $Y_2O_3$: $Nd^{3+}$(1at%) nanoparticles. (c) Particle size distribution of $Y_2O_3$: $Er^{3+}$(1at%), $Nd^{3+}$(1at%) and $Y_2O_3$:$Nd^{3+}$(1at%) nanoparticles.

**Fig. 4.** Measured XRF spectra of $Y_2O_3$: $Er^{3+}$(1at%), $Nd^{3+}$(1at%) and $Y_2O_3$: $Nd^{3+}$(1at%) nanopowder.

**Fig. 5.** The optical transmission spectra of $Y_2O_3$: $Er^{3+}$(1at%), $Nd^{3+}$(1at%) sintered pellet. Energy level transition corresponding to absorption in $Nd^{3+}$ (in black) and $Er^{3+}$ (in sky blue) are shown.

**Fig. 6.** (a) The infrared excitation spectra of both [$Y_2O_3$: $Er^{3+}$(1at%)$Nd^{3+}$(1at%) and $Y_2O_3$:$Nd^{3+}$(1at%)] transparent ceramics by monitoring emission at 1064 nm. The peaks are assigned corresponding to their energy transitions. (b) The PL emission spectra, under 822 nm excitation, of $Y_2O_3$:$Er^{3+}$(1at%)$Nd^{3+}$(1at%) and $Y_2O_3$:$Nd^{3+}$(1at%) samples in 420-1700 nm range. The peaks are assigned corresponding to their energy level transitions.

**Fig. 7.** Possible energy level diagram of $Nd^{3+}$ and $Er^{3+}$.

**Fig. 8.** Color coordinates of upconversion emission of $Y_2O_3$:$Er^{3+}$(1at%)$Nd^{3+}$(1at%) transparent ceramics under 822 nm excitation in 1931 CIE diagram.

**Fig. 9.** Near UV excitation spectra of both [$Y_2O_3$: $Er^{3+}$(1at%)$Nd^{3+}$(1at%) and $Y_2O_3$:$Nd^{3+}$(1at%)] transparent ceramics measured by monitoring emission at 1064 nm. The PL emission spectra, observed under 357 nm excitation of both samples in 400 -1500 nm range. The peaks are assigned corresponding to involved energy transitions in $Nd^{3+}$ and $Er^{3+}$ ions.

Nd$^{3+}$ Single-Doped, and Er$^{3+}$/Nd$^{3+}$ Codoped Tellurite Glasses for Mid-Infrared Laser Applications", *J. Am. Ceram. Soc.*, 94 (2011)1766.

**Figure 1:**

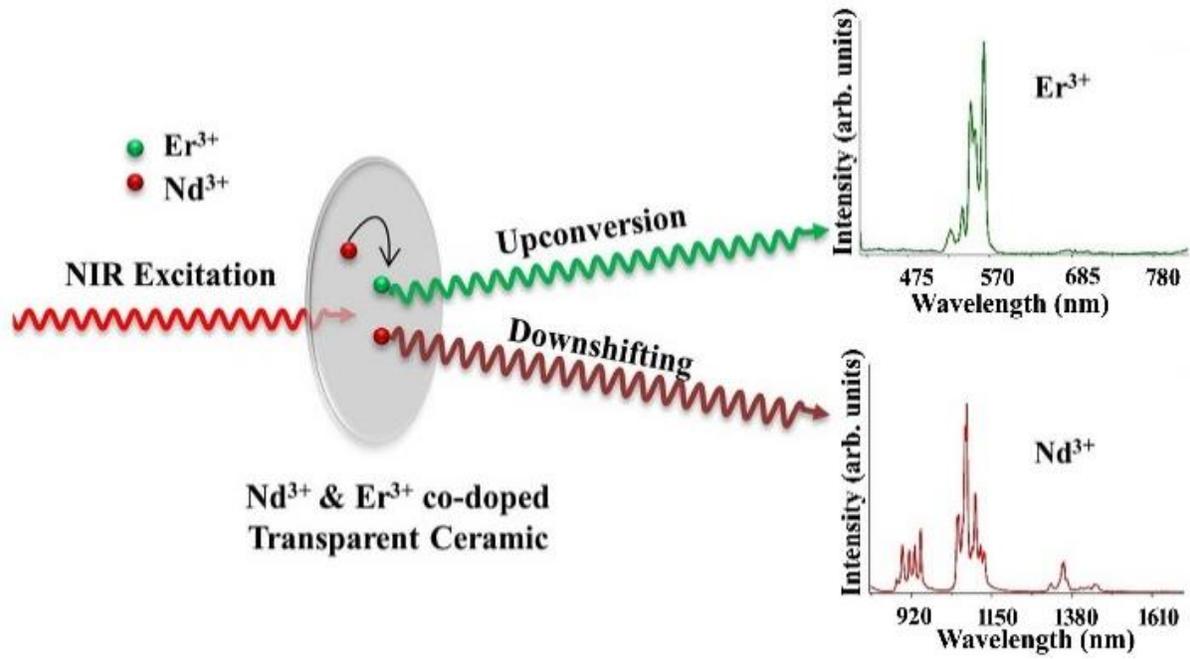

**Figure 2:**

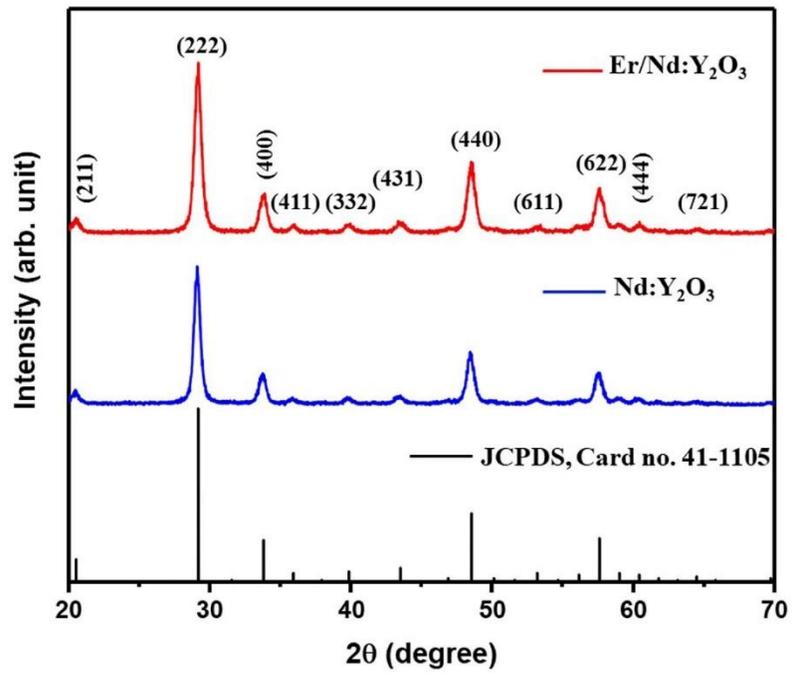

**Figure 3:**

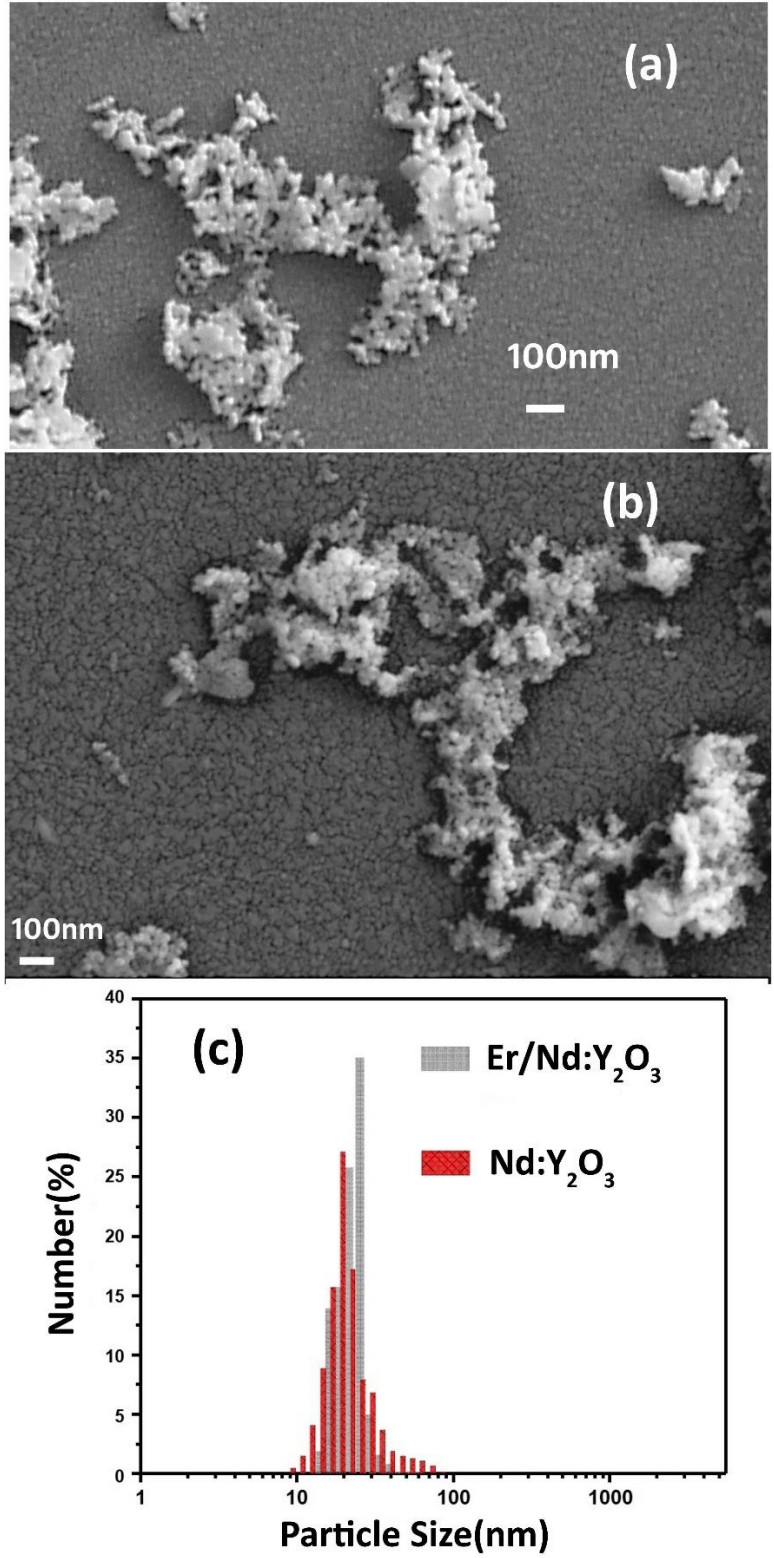

**Figure 4:**

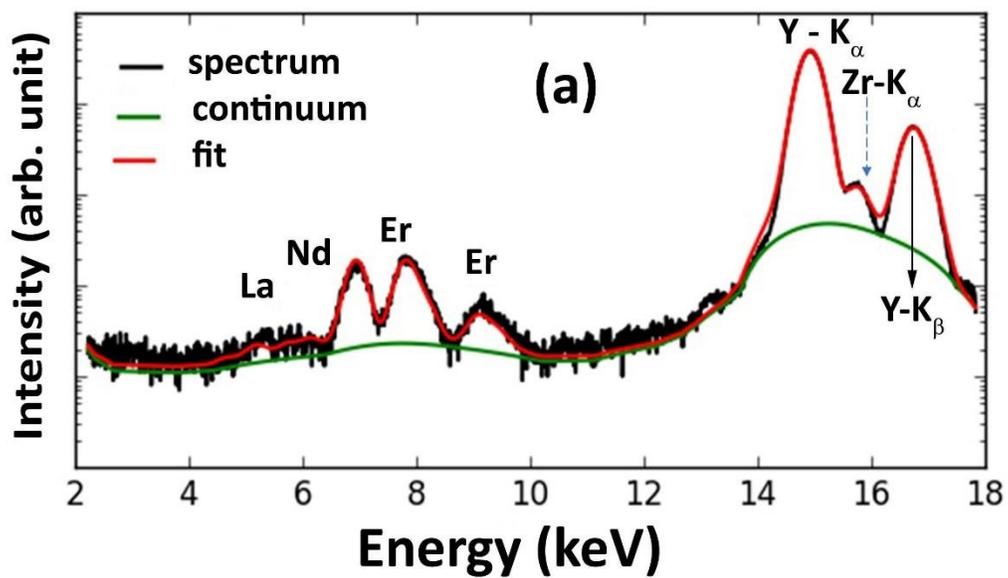
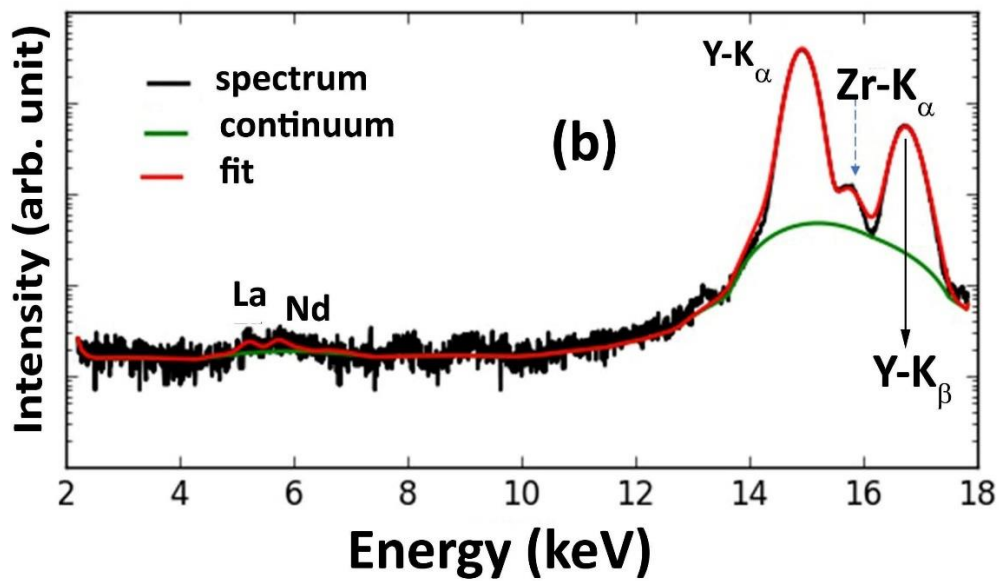

**Figure 5:**

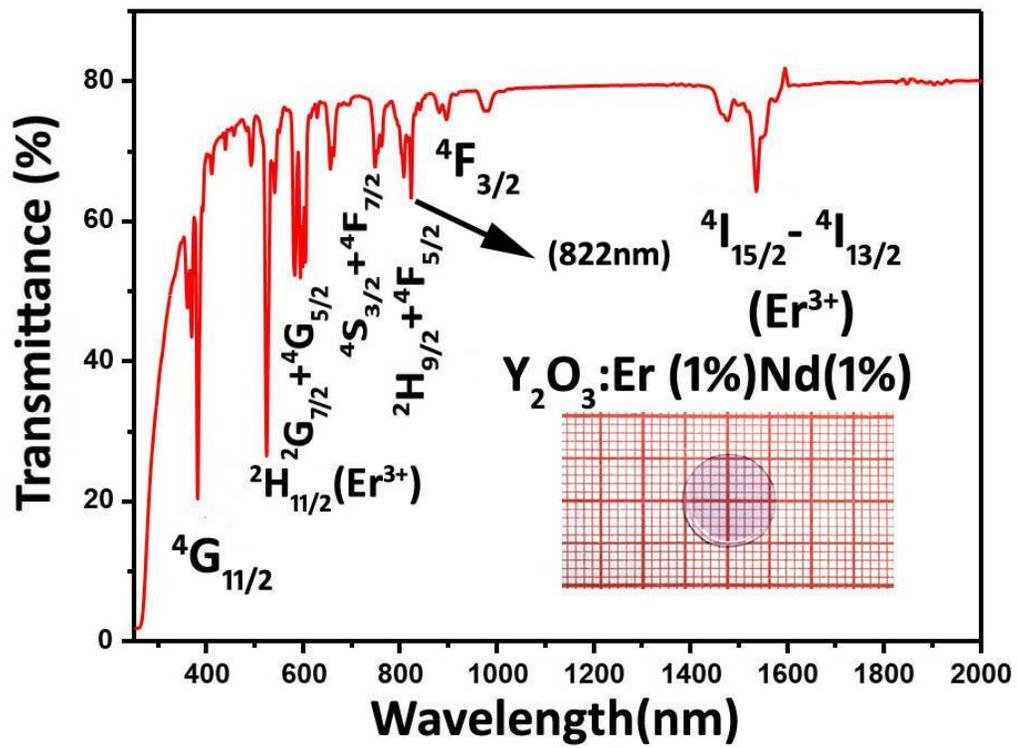

**Figure 6:**

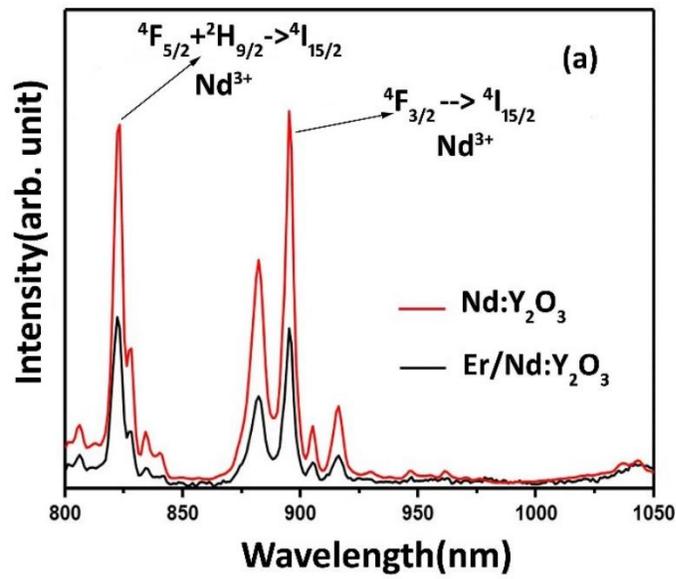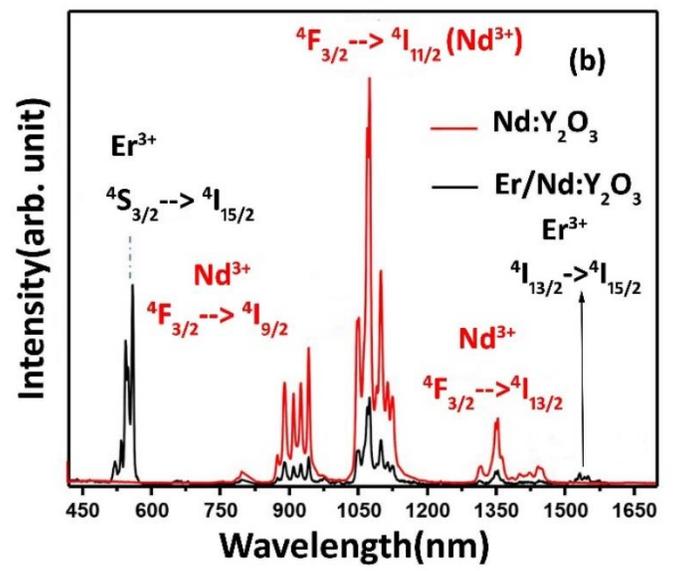

**Figure 7:**

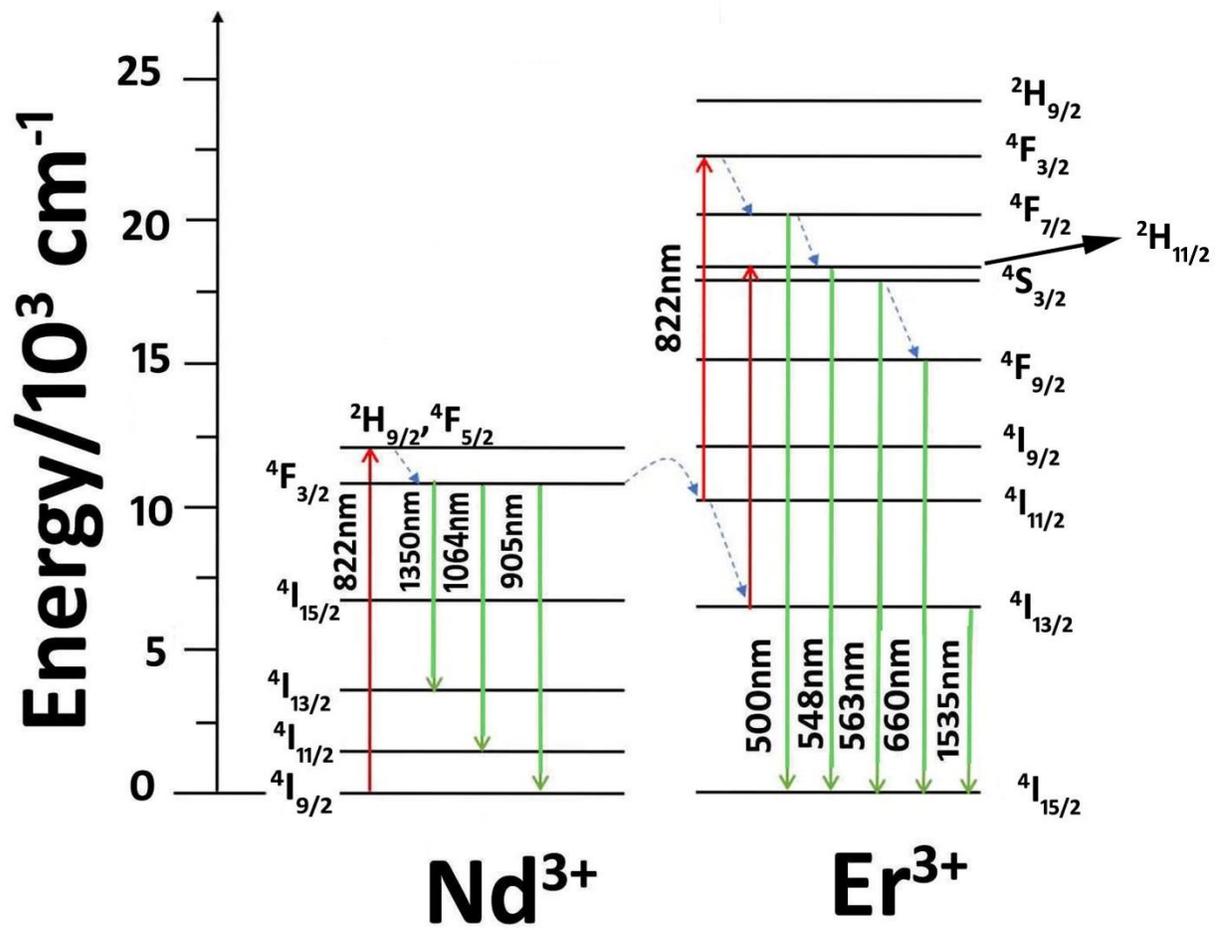

**Figure 8:**

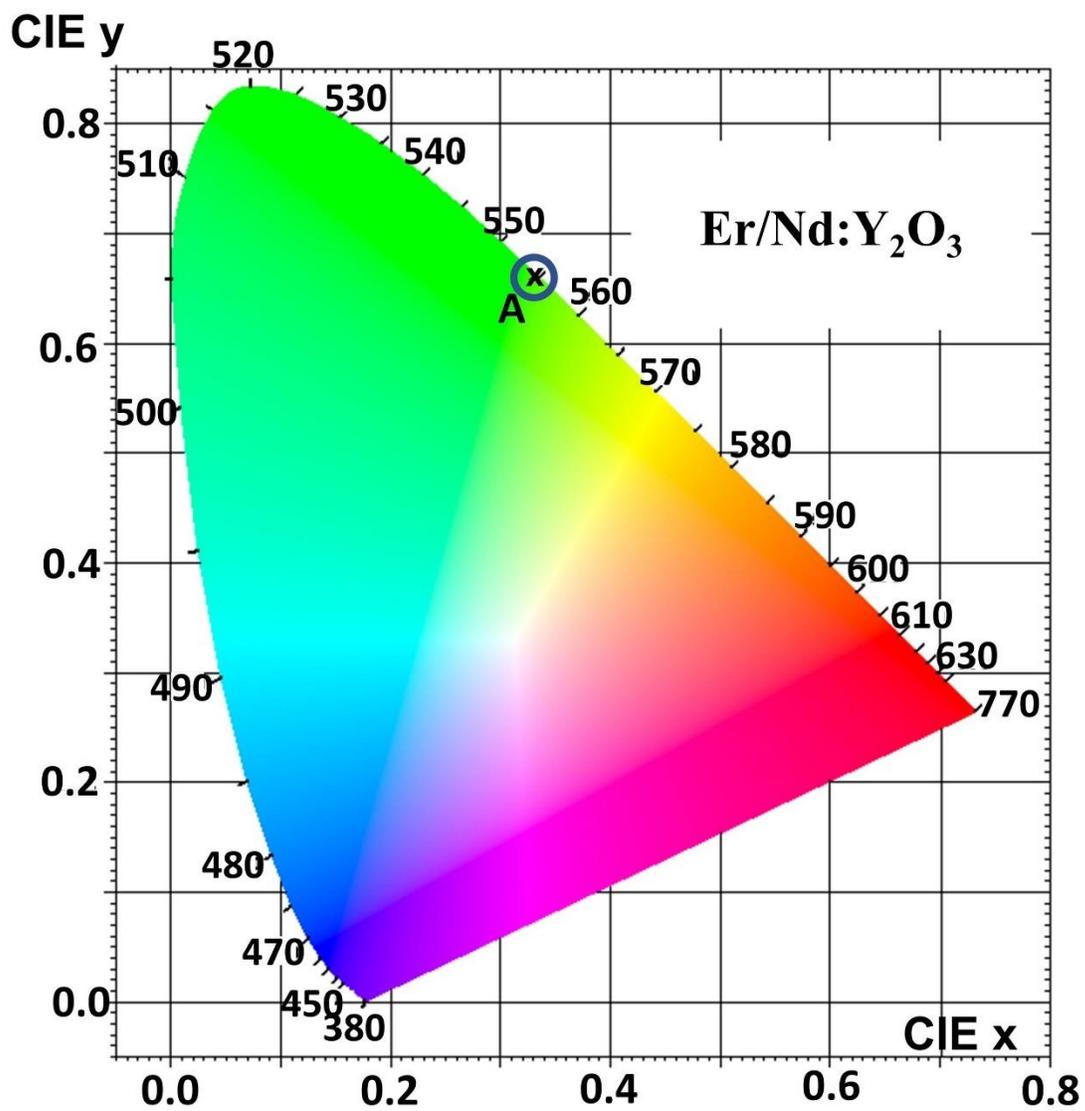

**Figure 9:**

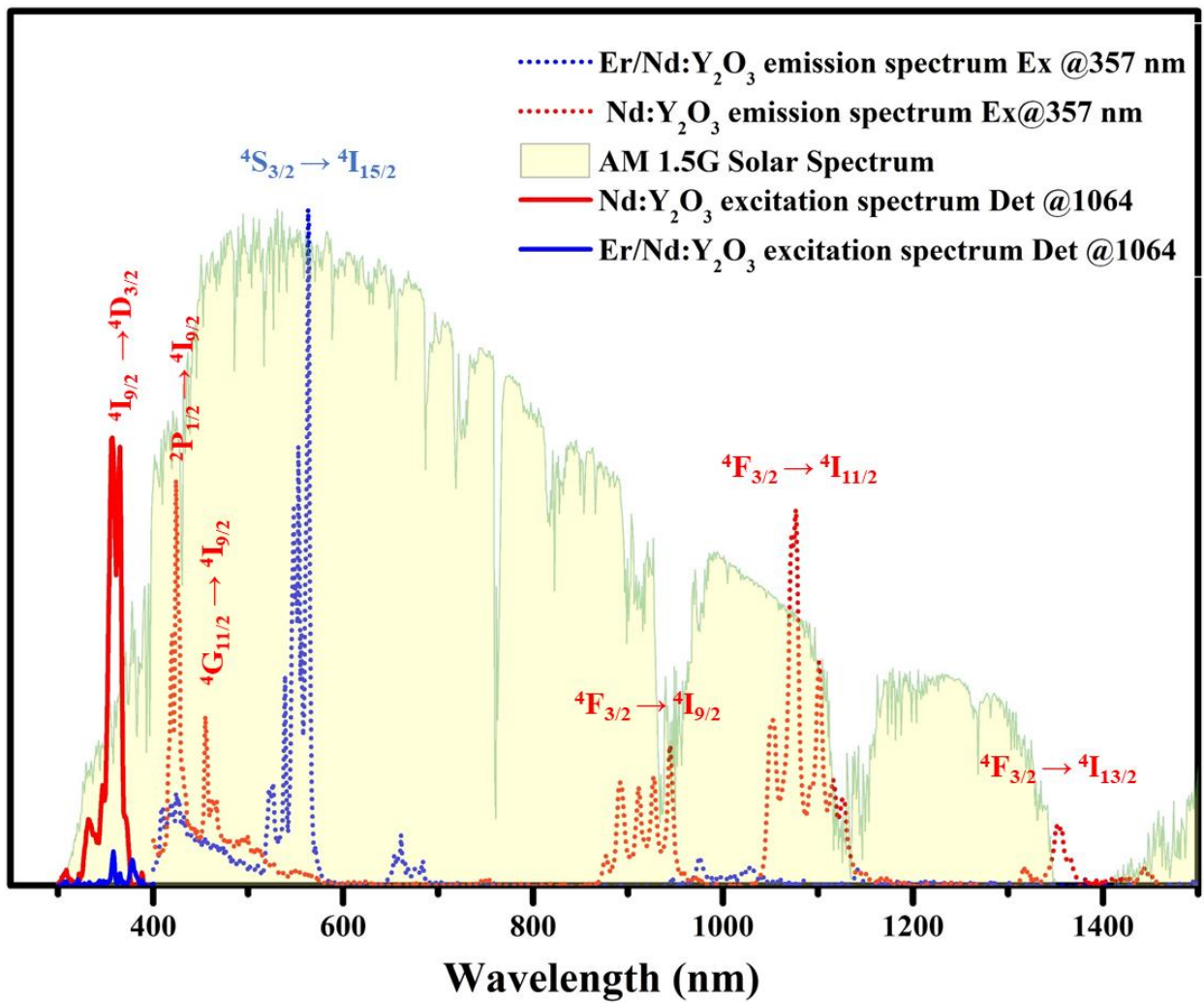